\documentclass[sigconf]{acmart}
\renewcommand\footnotetextcopyrightpermission[1]{} 
\usepackage{algorithm}
\usepackage{algpseudocode}

\usepackage{booktabs} 

\usepackage{url}
\usepackage{breakurl}
\hypersetup{draft} 

\setcopyright{rightsretained}
\usepackage{balance}

\newcommand{\method}{{\tt hood2vec}}

\acmDOI{10.475/123_4}

\acmISBN{123-4567-24-567/08/06}

\usepackage{subfigure}
\acmArticle{4}
\acmPrice{15.00}
\usepackage{xcolor}

\begin{document}
\title{\method: Identifying Similar Urban Areas Using Mobility Networks}

\author{Xin Liu}
\affiliation{%
  \institution{University of Pittsburgh}
  \city{Pittsburgh}
  \state{Pennsylvania}
}
\email{xil178@pitt.edu}

\author{Konstantinos Pelechrinis}
\affiliation{%
  \institution{University of Pittsburgh}
  \city{Pittsburgh}
  \state{Pennsylvania}
}
\email{kpele@pitt.edu}

\author{Alexandros Labrinidis}
\affiliation{%
  \institution{University of Pittsburgh}
  \city{Pittsburgh}
  \state{Pennsylvania}
}
\email{labrinid@cs.pitt.edu}






\renewcommand{\shortauthors}{B. Trovato et al.}

\begin{abstract}

Which area in NYC is the most {\em similar} to Lower East Side? 
What about the NoHo Arts District in Los Angeles? 
Traditionally this task utilizes information about the type of places located within the areas and some popularity/quality metric. 
We take a different approach. 
In particular, urban dwellers' time-variant mobility is a reflection of how they interact with their city over time. 
Hence, in this paper, we introduce an approach, namely {\method}, to identify the similarity between urban areas through learning a node embedding of the mobility network captured through Foursquare check-ins. 
We compare the pairwise similarities obtained from {\method} with the ones obtained from comparing the types of venues in the different areas. 
The low correlation between the two indicates that the mobility dynamics and the venue types potentially capture different aspects of similarity between urban areas. 

\end{abstract}

%
%


\maketitle

\section{Introduction}
\label{sec:intro}

Identifying similar areas in a city can facilitate dwellers and visitors exploring the city better. 
An intuitive approach for comparing urban areas and providing recommendations is to utilize information about the type of venues within an area.   
The types of venues within an area (zip code, neighborhood, block etc.) can be thought of as the ``signature'' of this area. 
However, there are several assumptions behind this consideration. 
For example, this implicitly assumes that all venues within the area are {\em active} through the whole day. 
For instance, while two areas can appear to have similar venues, they can be significantly different when introducing the temporal dimension (e.g., an area with ``lunch restaurants'' compared to one with ``dinner restaurants''). 
Furthermore, this static view does not include information on how urban dwellers interact with these areas, as captured through their movements between them. 

In this paper, we explore the mobility patterns of Foursquare users in the three US cities included in the Future Cities Challenge (FCC) dataset, namely, New York, Los Angeles and Chicago, borrowing analytical tools from the network science literature. 
In particular, we build {\method} that first designs a network between urban areas in a city using the FCC dataset and then obtains a vector representation of each area using a network embedding. 
We then use these representations to identify similar areas and make comparisons with other similarity metrics based purely on venue types and checkins. 
{\em Similar} urban areas will be represented by points (i.e., vector) closer together in the latent space identified by the network embedding. 
We then calculate the similarity between two urban areas using the (Euclidean) distance between the latent space points for the two areas. 

Existing literature has attempted to identify the functionality of urban areas, and consequently, cluster areas based on their functionality. 
Topic modeling is the dominant techniques in this line of research (e.g., \cite{cranshaw2010seeing,yuan2012discovering}). 
Other studies have attempted to identify similar areas across cities mainly using the type of activities recorded in the different areas of the different cities (e.g., \cite{le2015soho,fsq-similar}).
In the rest of the paper we formally describe our approach and present the results obtained from the three cities aforementioned. 

\section{{\method}: learning an urban area vector representation}
\label{sec:represent}

The FCC dataset provides information about the mobility patterns of Foursquare users. 
Each data point has the following tuple format: {\tt <start venue, end venue, trip year and month, trip period in a day, number of checkins>}. 
The {\em number of checkins} captures the number of times that the specific transitions were observed in the dataset. 
The dataset also provides information about the name, geographic coordinates and category for each venue. 

The majority of the transitions recorded in the dataset are observed only one time. 
In particular, 95\% of the transitions are observed less than 3 times. 
In order to avoid fitting the noise, we aggregate the transitions (movements) over a wider geographical scale. 
We also separate the movements according to the time period of movement occurrence according to the data - i.e., overnight (00:00 to 05:59), morning (06:00 to 09:59), midday (10:00 to 14:59), afternoon (15:00 to 18:59), night (19:00 to 23:59). 
Using MapQuest's Geocoding API\footnote{https://developer.mapquest.com/documentation/geocoding-api/} we obtain the zip code for each venue and we aggregate the movements at the zip code level (the wider scale). 
More specifically, we transform the original data to the following format per period: \textsf{<start zip code, end zip code, trip year and month, number of checkins>}. 
At zip code level, only 10\% of the movements have less than 2 observations. 
However, 20\% of the zip codes contain fewer than 10 venues and hence, we filter them out from our analysis. 
While this might sound a large number to ignore, the checkins within these zip codes cover only 0.5\% of the total checkins in the dataset. 

Finally, for each city $c \in \mathcal{C}$ = \{New York, Los Angeles, Chicago\} we define its directed urban flow network $\mathcal{G}_{c,p}$ per period $p \in \mathcal{P}$ = \{overnight, morning, midday, afternoon, night\} at the zip-code level as follows: 
$\mathcal{G}_{c,p}  = (\mathcal{U}, \mathcal{E})$, where the set of nodes $\mathcal{U}$ is the set of zip code areas in city $c$. 
A directed edge $e_{ij} \in \mathcal{E}$ exists between two zip codes $u_i, u_j \in \mathcal{U}$ if there has been observed at least one movement from a venue in $u_i$ to a venue in $u_j$ during period $p$. 
We also annotate every edge $e_{ij}$ with a weight $w(e_{ij})$, which captures the number of checkins of such movements observed. 

We would like to note here that while we have chosen the zip codes as our unit, one can define an urban area differently. 
For instance, one can use the notion of {\em neighborhoods} that can include several zip codes, or the census tracts, or any other definition of neighborhood \cite{cranshaw2012livehoods}. 

\subsection{Vector Representation by {\tt node2vec}}

In order to obtain a vector representation for the nodes of $\mathcal{G}_{c,p}$, i.e., the zip codes at $c$ in period $p$, we will rely on learning a network embedding.   
There are several ways to learn a node embedding for a network but in this work we make use of {\tt node2vec} \cite{grover2016node2vec}. 
%
%
Briefly, {\tt node2vec} utilizes second order random walks to learn a vector representation for the network nodes that optimizes a neighborhood preserving objective function.  
The framework is flexible enough to accommodate various definitions of network neighborhood and facilitate the projections of the network nodes in the latent space according to different {\em similarity} definitions. 
Given that we are interested in the structural equivalence of the urban areas, we pick the parameters of {\tt node2vec} accordingly ($p=1$ and $q=2$ \cite{grover2016node2vec}). 
We also utilize 1,000 random walks for the sampling process, while  
we set the dimensionality of the latent space to $d=10$. This is consistent with the dimensionality of another vector representation to be introduced in Section \ref{sec:venue-rep}.
{\tt node2vec} finally provides us with a vector $\mathbf{v}_i \in \mathbb{R}^d, \forall u_i \in \mathcal{U}$, that we can then use to identify the similarity between two urban areas. 


\subsection{Vector Representation utilizing Venue Categories}
\label{sec:venue-rep}

As alluded to above, a straightforward way to define the similarity between two areas is to compare the distribution of the type of venues they host. 
More specifically we can define a vector $\mathbf{z}_i$ for each urban area node, such that its $k^{th}$ element $z_{ik} = \dfrac{n_{ik}}{N_i}$, where $n_{ik}$ is the number of venues of type $k$ within area $i$ and $N_i$ is the total number of venues within $i$.  
For defining vectors $\mathbf{z}_i$ we use the 10 top-level venue categories in Foursquare (thus, $\mathbf{z}_i \in \mathbb{R}^{10}$)
: Arts \& Entertainment, College \& University, Event, Food, Nightlife Spot, Outdoors \& Recreation, Professional \& Other Places, Residence, Shop \& Service, Travel \& Transport. 
Similar to {\method}, we can now define the similarity between two urban areas $i$ and $j$ using the distance between vectors $\mathbf{z}_i$ and $\mathbf{z}_j$. 

Similar to number of venues, the number of checkins in venues of different types can also be used as the vector representation of an urban area. 
In particular, we define a vector $\mathbf{z}_i^{\text{check}}$ for each urban area node, such that its $k^{th}$ element $z_{ik}^{\text{check}} = \dfrac{n_{ik}}{C_i}$, where $c_{ik}$ is the number of checkins of venues of type $k$ within area $i$ and $C_i$ is the total number of checkins of venues within $i$. We follow the same 10 top-level venue categories for $\mathbf{z}_i^{\text{check}}$ (i.e., $\mathbf{z}_i^{\text{check}} \in \mathbb{R}^{10}$). Then we can also define the similarity between two areas $i$ and $j$ by the distance between vectors $\mathbf{z}_i^{\text{check}}$ and $\mathbf{z}_j^{\text{check}}$. 



\section{Urban Area Similarity}
\label{sec:methods}

One of the questions is which urban area representation should we use? 
Do they even provide us with a different view of the similarity between two areas? 
In order to explore this we will calculate the pairwise similarities using the network embedding learnt from {\method} and compare them with the corresponding pairwise similarities obtained from a simple venue-based representation of urban areas (see Section \ref{sec:venue-rep}). 
Formally, the similarity of two areas $i$ and $j$, with vector representations $\mathbf{x}_i$ and $\mathbf{x}_j$ respectively, is defined as: 

\begin{equation}
    \sigma_{ij} = \exp({-{\tt dist}_s(\mathbf{x}_i,\mathbf{x}_j)})
    \label{eq:sim}
\end{equation}
where ${\tt dist}(\mathbf{x}_i,\mathbf{x}_j)$ is the  (Euclidean) distance between the representations of $i$ and $j$. 

We can now examine whether different representations for the urban areas provide different views for their similarity. 
In particular, if $\sigma_{ij}$ and $\sigma^{'}_{ij}$ are the similarities between areas $i$ and $j$ using different vector representations, their Pearson correlation coefficient $\rho_{\sigma,\sigma^{'}}$ will be high if the two representations provide similar information, and low otherwise. 
We can further compare in the same way the similarity of two areas for the same vector representation over different time periods. 

\section{Experiments and Results}
\label{sec:results}

In this section, we will present the results of our analysis and compare the pairwise similarities obtained from {\method} and a simple venue category-based representation. 

\subsection{Movement and Venue Categories}


We calculate the correlation between two representations, $\mathbf{v}$ and $\mathbf{z}$, (by the method in Section \ref{sec:methods}) in three cities: New York City, Los Angeles and Chicago. 
There is a total of 141 zip codes $u_i$ (9870 pairs) in New York city, 111 zip codes (6105 pairs) in Los Angeles, and, 59 zip codes (1711 pairs) in Chicago. 
We further extend our comparisons to each time period provided in the data. 
The results are presented in Table \ref{tab:rsquared}. 
Note that we use the following notation for the five time periods - O: overnight; MO: morning; MI: midday; A: afternoon; N: night. 
As we can see all the correlations are positive, albeit, small, pointing to the two representations capturing different types of information. 
We also calculate the correlation between $\mathbf{z}$ and $\mathbf{z}^{\text{check}}$ in three cities. The correlations for these three cities are 0.839, 0.930, 0.936, respectively. I.e., the representations of venue category based on number of venues and checkins are highly correlated. This indicates low correlation of representations between {\method} and checkin-based venue category.

\begin{table}[]\centering
\begin{tabular}{c c c c c c}
\toprule
\textbf{Period} & \textbf{O} & \textbf{MO} & \textbf{MI} & \textbf{A} & \textbf{N} \\ 
\midrule
New York & 0.116\textsuperscript{***} & 0.152\textsuperscript{***} & 0.147\textsuperscript{***} & 0.152\textsuperscript{***} & 0.144\textsuperscript{***}\\
Los Angeles & 0.184\textsuperscript{***} & 0.290\textsuperscript{***} & 0.229\textsuperscript{***} & 0.219\textsuperscript{***} & 0.142\textsuperscript{***}\\
Chicago & 0.284\textsuperscript{***} & 0.316\textsuperscript{***} & 0.327\textsuperscript{***} & 0.336\textsuperscript{***} & 0.323\textsuperscript{***}\\
\bottomrule
\addlinespace[1ex]
\multicolumn{3}{l}{\textsuperscript{***}$p<0.01$, 
  \textsuperscript{**}$p<0.05$, 
  \textsuperscript{*}$p<0.1$}
\end{tabular}
\caption{Correlation between movement and category representations.}
\label{tab:rsquared}
\vspace{-0.30in}
\end{table}

\begin{figure}
\includegraphics[width=6.3cm]{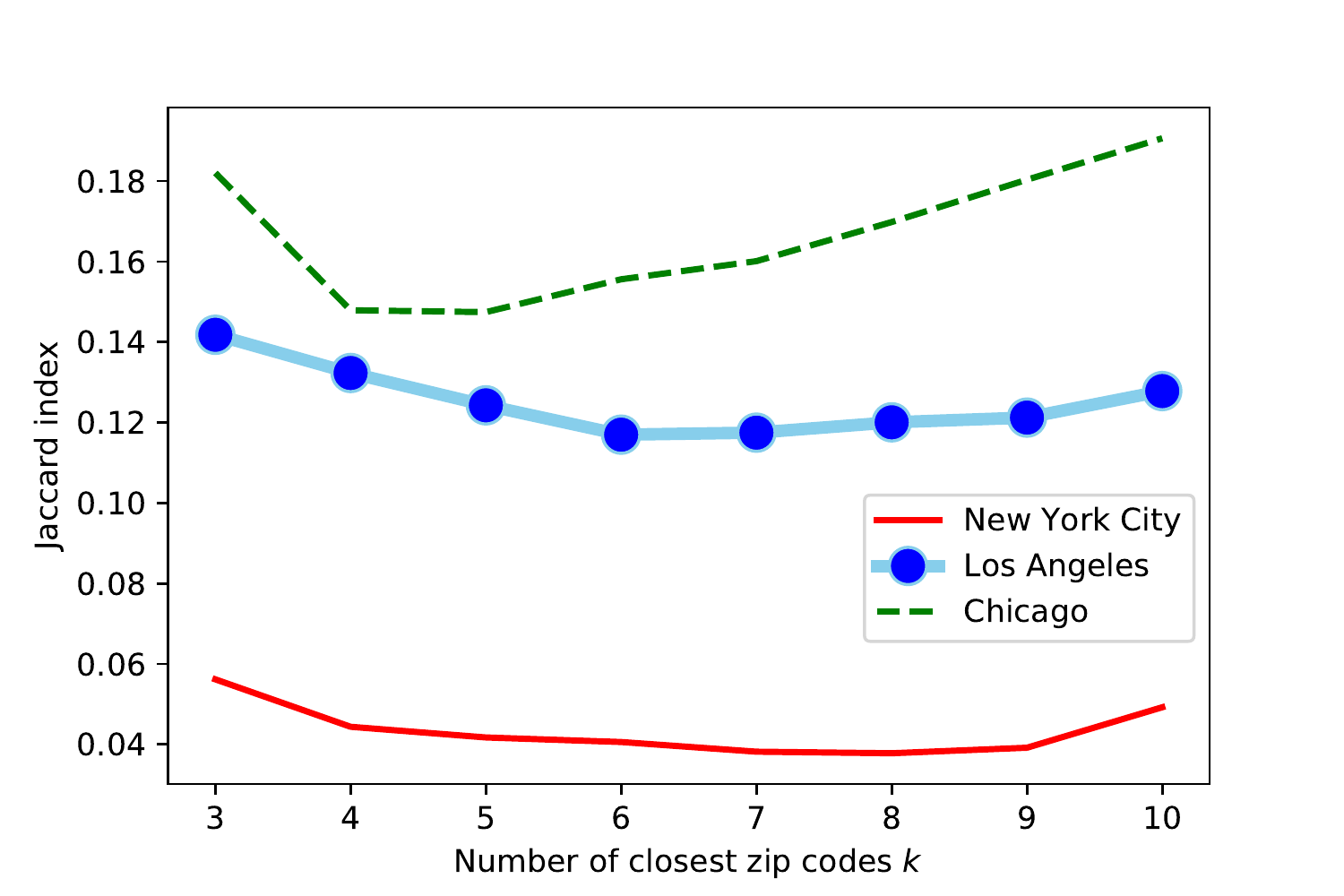}
\caption{Average Jaccard index as function of number of closest neighbors $k$.}\label{fig:J_Index}
\end{figure}

We further inspect the relationship between the two approaches from the perspective of the top-k neighbors for each zip codes. 
In particular, for each zip code $u_i$ we find the $k = 5$ closest zip codes to $i$ based on their {\method} representation ($\mathbf{v}$), $\mathcal{N}_{5,i,\method}$. 
Similarly, we calculate the top-5 neighbors of zip code $u_i$ based on their venue category representation ($\mathbf{z}$), $\mathcal{N}_{5,i,cat}$. 
We then calculate the Jaccard index of the two sets: 

\begin{equation}
    J(\mathcal{N}_{5,i,\method},\mathcal{N}_{5,i,cat}) = \dfrac{|\mathcal{N}_{5,i,\method} \cap \mathcal{N}_{5,i,cat}|}{|\mathcal{N}_{5,i,\method} \cup \mathcal{N}_{5,i,cat}|}
   \label{eq:jaccard} 
\end{equation}
Table \ref{tab:overlap} presents the average Jaccard index for every city and time period. 
Furthermore, Figure \ref{fig:J_Index} presents the Jaccard index as a function of the number of neighbors $k$ considered for every city, averaged over different time periods and zip codes.
As one might have expected from the earlier results presented, in general, under different $k$, there are few shared neighbors when using the two different representations for the zip codes.
This strengthens our hypothesis that these two types of representations capture different information for the areas. 

\begin{figure}
\includegraphics[width=5.0cm]{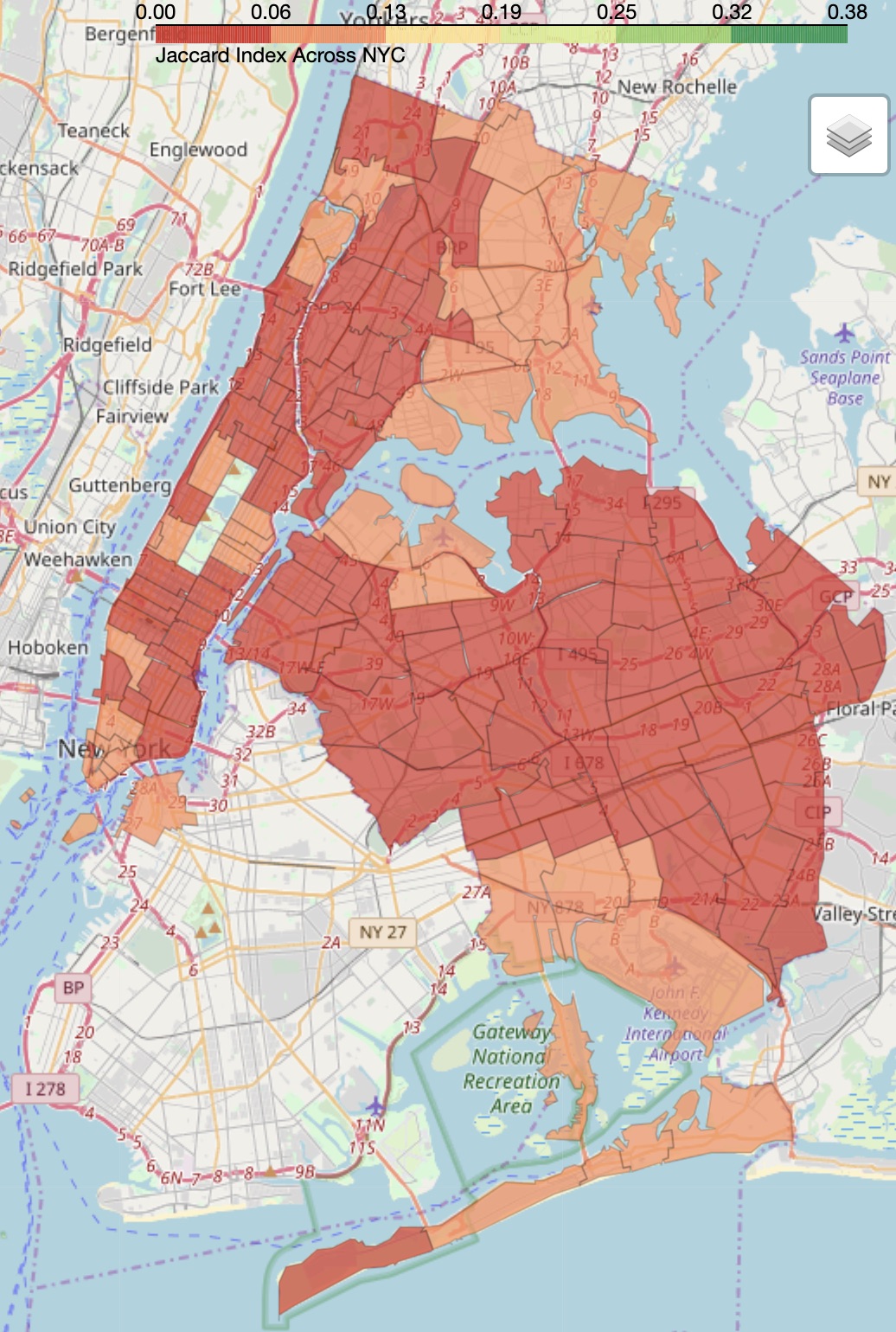}
\caption{Average Jaccard index over the different time-periods for the zip codes New York City.}\label{fig:NYC}
\end{figure}

\begin{figure}
\includegraphics[width=5.0cm]{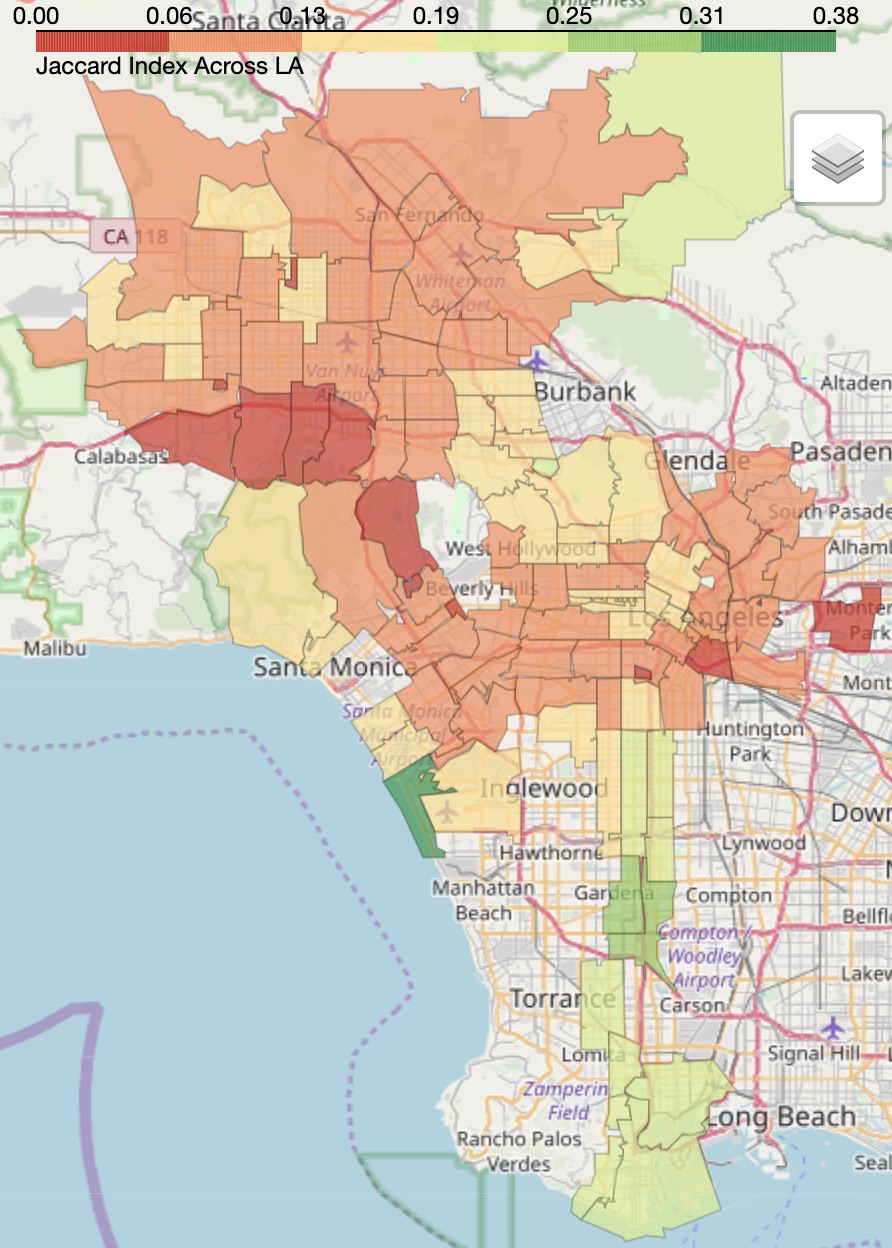}
\caption{Average Jaccard index over the different time-periods for the zip codes in Los Angeles.}\label{fig:LA}
\end{figure}

\begin{figure}
\includegraphics[width=5.0cm]{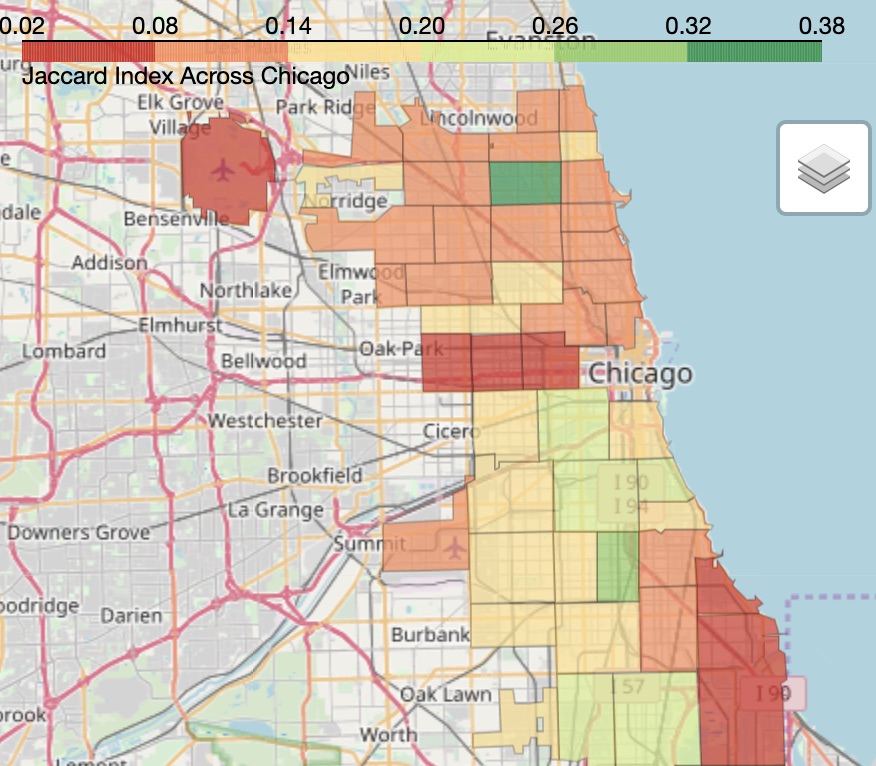}
\caption{Average Jaccard index over the different time-periods for the zip codes in Chicago.}\label{fig:CHI}
\end{figure}

Moreover, Figures \ref{fig:NYC}-\ref{fig:CHI} illustrate the Jaccard index for every zip code per city, averaged over the different time periods. 
As we can see most of the zip codes in all cities have a fairy low Jaccard index.  
New York City's zip codes exhibit overall lower Jaccard index compared to Chicago and Los Angeles (in accordance to the results in Table \ref{tab:rsquared}, \ref{tab:overlap}).  
Zip codes with high Jaccard index are essentially urban areas for which the two different representations examined  identify a high overlap on areas similar to them. 
This happens to a larger extend in Los Angeles and Chicago compared to New York City. 
This can potentially be due to (a) the compact nature of NYC that allows people to explore several different areas and hence, geographically remote zip codes are close in the {\method} latent space, and/or, (b) the different geographic distribution of venues in the three different cities. 
More specifically, the compact nature may cause venues in New York city more evenly distributed, since they are easily accessible by dwellers. In contrast, scattered nature of Los Angeles may lead to biased venue distribution due to various accessibility of different regions; this could be the reason for slightly high Jaccard indices in some areas. Chicago has  fewer zip code areas such that an area can has higher probability of sharing the same closest area(s) in two representations; this can cause slightly high Jaccard indices in some areas.
Nevertheless, regardless of the reasons for the differences across the cities examined, in all cases the Jaccard index does not go beyond 0.4. 
Simply put, there is no zip-code in these three cities, for which the overlap between the top-5 neighbors identified by {\method} and a simple venue-based vector representation is more than 40\%, supporting our hypothesis that these two different approaches capture different information with respect to the similarity of the areas. 



\begin{table}[]\centering
\begin{tabular}{c c c c c c}
\toprule
\textbf{Period} & \textbf{O} & \textbf{MO} & \textbf{MI} & \textbf{A} & \textbf{N} \\ 
\midrule
New York City & 0.036 & 0.065 & 0.062 & 0.029 & 0.016 \\
Los Angeles & 0.098 & 0.254 & 0.150 & 0.104 & 0.015\\
Chicago & 0.139 & 0.136 & 0.170 & 0.164 & 0.129\\
\bottomrule
\addlinespace[1ex]
\end{tabular}
\caption{Jaccard index ($k=5$) for the three cities for the different time periods averaged over the corresponding zip codes.}
\label{tab:overlap}
\vspace{-0.30in}
\end{table}

\subsection{{\method} representation across time}

\begin{figure}
\includegraphics[width=5.3cm]{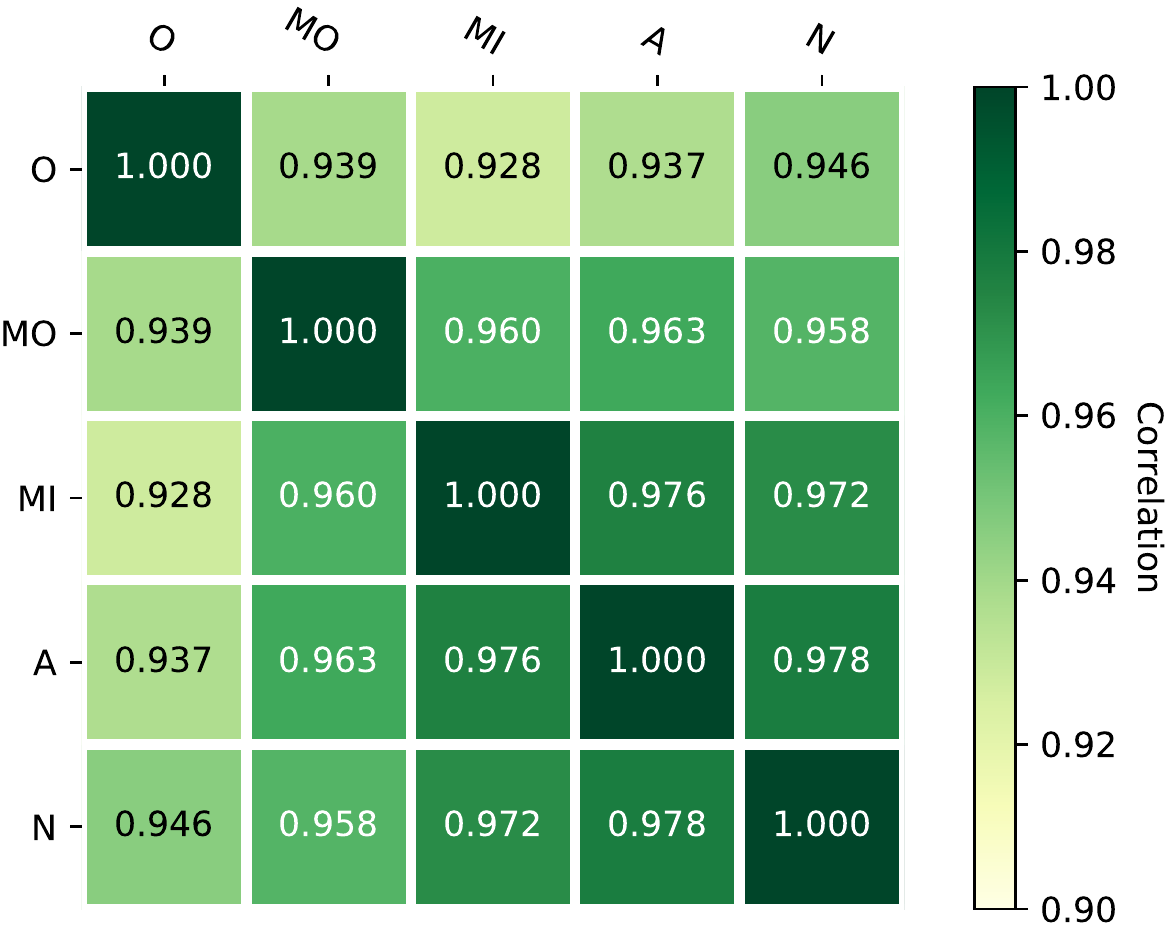}
\caption{Correlation among representations of different periods in New York City.}\label{fig:NYC_period}
\end{figure}

\begin{figure}
\includegraphics[width=5.3cm]{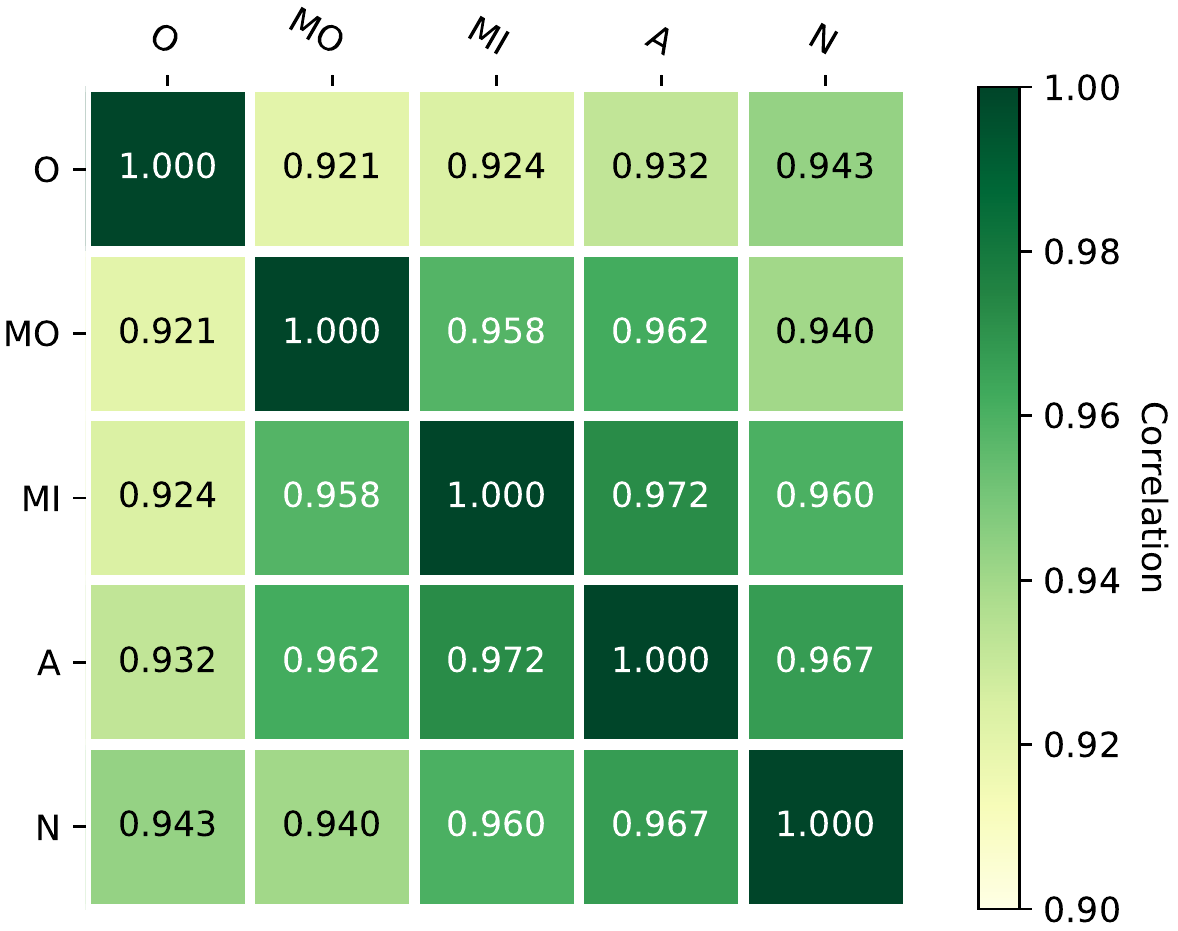}
\caption{Correlation among representations of different periods in Los Angeles.}\label{fig:LA_period}
\end{figure}

\begin{figure}
\includegraphics[width=5.3cm]{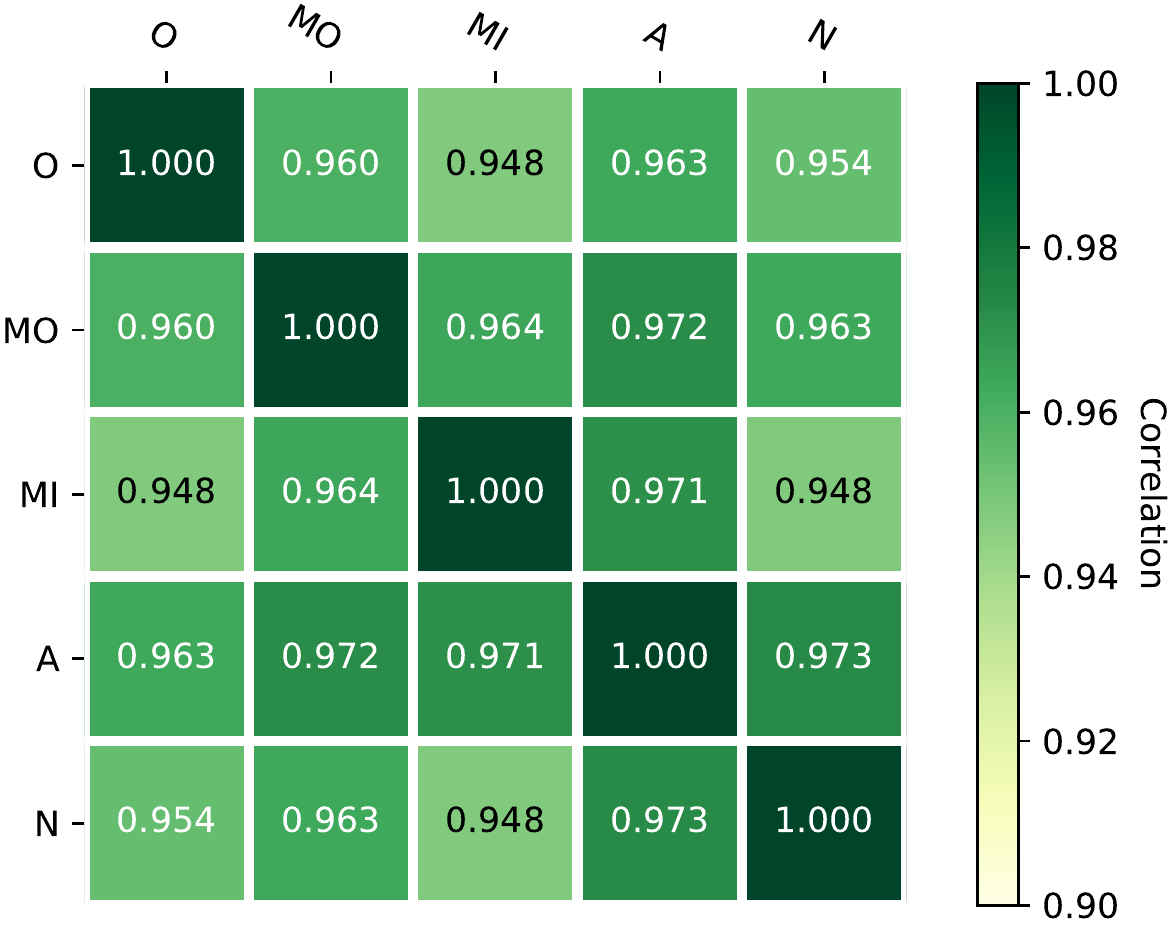}
\caption{Correlation among representations of different periods in Chicago.}\label{fig:CHI_period}
\vspace{-0.2in}
\end{figure}

We further explore how the representation obtained for a zip code through {\method} changes over time (i.e., over the different time-periods in the dataset). 
%
Let us assume the two periods $p_1$ and $p_2$, and the corresponding {\method} representation vectors $\mathbf{v}_{p_1}$ and $\mathbf{v}_{p_2}$ respectively. 
Then following similar steps as the ones described in Section \ref{sec:methods}, we can obtain the pairwise correlation of the  between periods $p_1$ and $p_2$ for the same city.
The correlations of each city are shown in Fig. \ref{fig:NYC_period}, \ref{fig:LA_period}, \ref{fig:CHI_period}.

One can observe that for these three cities, the correlations between any pair of periods are very high, all over 0.9. This means that, the patterns of movements are similar regardless of the time of a day (based on the {\method} representation). 
Since New York City and Chicago are more geographically compact, it is easier for dwellers to move within the city for any purpose at any time. This could be the reason that the overall movement patterns within a day are similar. Los Angeles is geographically scattered, which limits the convenience of movements; dwellers tend to move within nearby areas at any time of the day. This may cause similar movement patterns of all day.
The interested reader can explore the different urban area representations at: \url{http://www.pitt.edu/~xil178/hood2vec.html}
\section{Conclusions}
\label{sec:conclusions}

In this paper, we propose {\method} to identify the similarity between urban areas through learning a node embedding of the mobility network captured through Foursquare check-ins. 
We compare the pairwise similarities obtained from {\method} with the ones obtained from comparing the types of venues in the different areas. 
The low correlation between the two indicates that the mobility dynamics and the venue types potentially capture different aspects of similarity between urban areas. 

{\bf Acknowledgments: }This work is part of the PittSmartLiving project (\url{https://pittsmartliving. org/}) which is supported in part by National Science Foundation Award CNS-1739413.

\bibliographystyle{ACM-Reference-Format}
\bibliography{sample-bibliography}
\balance

\end{document}